\documentclass[a4paper,11pt]{article}
\usepackage{pos}
\usepackage{subfigure}
\usepackage{aas_macros}

\title{VLBI Probes of Jet Physics in Neutrino-Candidate Blazars}
 \ShortTitle{VLBI Probes of Jet Physics in Neutrino-Candidate Blazars}

\author*[a,b]{F.~Eppel}
\author[a]{M.~Kadler}
\author[b]{E.~Ros}
\author[b,a]{P.~Benke}
\author[c]{M.~Giroletti}
\author[a,b]{J.~He{\ss}d{\"o}rfer}
\author[d]{F.~McBride}
\author[a,b]{F.~Rösch}

\affiliation[a]{Julius-Maximilians-Universit\"at W\"urzburg, Physikalisches Institut, Lehrstuhl für Astronomie, Emil-Fischer-Stra{\ss}e 31, 97074 W\"urzburg, Germany; 
$^b$\,Max-Planck-Institut für Radioastronomie, Auf dem Hügel 69, 53121, Bonn, Germany; 
$^c$\,INAF-Istituto di Radioastronomia, Bologna, Via Gobetti 101, 40129, Bologna, Italy;
$^d$\,Department of Physics and Astronomy, Bowdoin College, Brunswick, Maine 04011, USA;
}

\emailAdd{florian@eppel.space}

\abstract{In recent years, evidence has accumulated that some high-energy cosmic neutrinos can be associated with blazars. The strongest evidence for an individual association was found in the case of the blazar TXS\,0506+056 in 2017. In July 2019, 
another
track-like neutrino event (IC190730A) was found spatially coincident with the well-known bright flat-spectrum radio quasar PKS\,1502+106. PKS\,1502+106 was not found to be in a particularly elevated gamma-ray state, but exhibited a remarkably
bright radio outburst at the time of the neutrino detection, similar to TXS\,0506+056. We have performed a multi-frequency VLBI study from 15\,GHz up to 86\,GHz on TXS\,0506+056, PKS\,1502+106 and one additional neutrino-candidate blazar (PKS\,0215+015) to study the radio structure of neutrino candidate blazars in 
response
to their neutrino association. 
We have obtained target of opportunity observations with the VLBA for all three sources within $\sim$1\,month from their associated neutrino events and 
are performing
multi-epoch studies of the jet kinematics at 15\,GHz as part of the MOJAVE program. Here, we present first results on TXS\,0506+056 at 86\,GHz and one additional 43\,GHz image obtained 27 days after IC170922A, closer in time to the neutrino event than previously published images. 
We also give an overview about our recent work on PKS\,1502+106 and PKS\,0215+015.
}

\ConferenceLogo{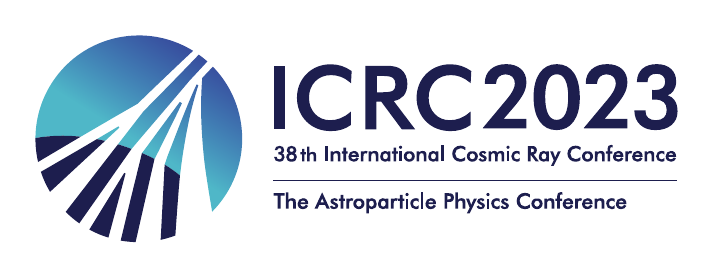}

\FullConference{%
38th International Cosmic Ray Conference (ICRC2023)\\
  26 July - 3 August, 2023\\
  Nagoya, Japan}

\begin{document}
\maketitle

\section{Introduction}
Gamma-ray emitting blazars have long been predicted to contribute significantly to the production of very-high-energy neutrinos above $\sim100$\,TeV \citep[e.g.,][]{Mannheim1995}. In the past years, observational evidence has started piling up, confirming that some high-energy cosmic neutrinos detected by the IceCube observatory can be associated with blazars. 
In addition,
several statistical studies suggest a connection between (radio-flaring) blazars and neutrino emission \citep[e.g.,][]{Plavin2020,Plavin2021,Hovatta2021,Buson2022,Buson2023, Krauss2018}.
However, recent studies of IceCube data have shown that blazars might only account for a small fraction of the diffuse neutrino flux across the whole energy range from a few TeV up to PeV energies \citep{IceCube2023}, which highlights the importance of studying the most likely individual associations.
In presence of the strong atmospheric background and high positional uncertainty of current neutrino observatories, multi-wavelength (MWL) information has proven to be crucial to identify and investigate the most intriguing individual cases of blazar-neutrino associations \citep{Kadler2016,IceCube2018a,IceCube2018b}.
The most significant blazar-neutrino association to date was found between the gamma-ray blazar TXS\,0506+056 and the neutrino event IC170922A in September 2017 (GCN Circular \#21916). Due to the good positional accuracy of this event and the variability of the gamma-ray emission, the significance of the association was found to be on the order of 3\,$\sigma$ \citep{IceCube2018b}. 
Moreover, the blazar was found to be in a major long-lasting gamma-ray and radio outburst at the time of the neutrino detection. An independent second analysis found a 3.5~$\sigma$ significant association between TXS\,0506+056 and a $\sim$160 day long flare of lower-energy neutrinos in 2014/15 \citep{IceCube2018a}. Surprisingly, this previous neutrino flare was not associated with enhanced activity in the gamma-rays in the \textsl{Fermi}/LAT band. Based on several modelling studies, it was found that the bulk of the LAT-photons received during the neutrino flare could not have been produced by the same mechanism as the IceCube-neutrinos \citep{Reimer2019,Rodrigues2019}.
It 
appears
that blazars can produce neutrinos both in bright (IC170922A),
and in faint (2014/15 ’flare’) gamma-ray states
via different processes.
Before mid 2019, no other bright blazar could be associated with high confidence with a
high-energy track-like muon neutrino event. On July 30, 2019, the track-like neutrino event IC190730A with high probability of being of astrophysical origin was found by IceCube. It spatially coincides within 0.31\,degrees with the well-known bright flat-spectrum radio quasar (FSRQ) PKS 1502+106 \citep[]{Karamanavis2016}. Unlike the TXS\,0506+056 association, this source was not found to be in a particularly high gamma-ray state at the time of the IceCube alert. 
However, PKS\,1502+106 had been exhibiting a bright radio outburst at the time of the neutrino event \citep{Kiehlmann2019}, similar to the case of TXS 0506+056. This substantially increases the probability of a 
real
physical association and 
raises
interest in the properties of the radio outbursts of neutrino-associated blazars.
More
recently, in February 2022, we have found the flat-spectrum radio quasar PKS\,0215+015 in a remarkably bright radio state in temporal and positional coincidence with the IceCube event IC220225A \citep{Eppel2023, Plavin2022}. The publicly available \textsl{Fermi}/LAT light curve reveals a simultaneous strong gamma-ray outburst, similar to the case of TXS\,0506+056 which makes this association very intriguing.

We have obtained target of opportunity (ToO) observations with the Very Long Baseline Array (VLBA) from 15\,GHz to 86\,GHz to 
probe
the parsec scale jet structure of TXS\,0506+056, PKS\,1502+106 and PKS\,0215+015 within $\sim$1\,month from their associated neutrino events. 
Very-long-baseline interferometry (VLBI) 
is the only way
to investigate the pc-scale jet structure down to the jet base 
with
high angular resolution in these blazars. 
One possible model for neutrino production \citep{Tavecchio2015} 
suggests
that the high-energy protons of the fast inner spine of the blazar jet interact with soft target photons of the sheath, a slower jet layer surrounding the spine, leading to high-energy $\gamma$-ray and neutrino emission. Other models suggest that neutrino production could happen closer to the jet base, e.g., in standing recollimation shocks \citep[e.g.,][]{Kalashev2022}. In principle, both 
features (spine-sheath structure and standing recollimation shocks) can be revealed by VLBI observations, 
but 
the compact structure of blazars requires high angular resolution observations to do this successfully. 
Our main goal is to search for these characteristic signatures of neutrino production in the jets of these blazars in total intensity and polarization. Here, we present first results on TXS\,0506+056 at 86\,GHz and one additional 43\,GHz image, closer in time to the neutrino event than previously published \citep{Ros2020}.  
We also
give an overview 
of
our recent work on PKS\,1502+106 and PKS\,0215+015.

\section{Observations \& Analysis}

We have observed TXS\,0506+056, PKS\,1502+106 and PKS\,0215+015 with the VLBA within $\sim$1 month of their associated neutrino events IC170922A, IC190730A and IC220225A, respectively. In the case of TXS\,0506+056, the source was observed on Oct 19, 2017 and Nov 10, 2017 at 15, 22, 43 and 86\,GHz, and on May 4, 2018 at 43 and 86\,GHz. Results for the Nov 2017 and May 2018 epoch at 43\,GHz were previously published \citep{Ros2020}. 
Here, we present the analysis of the 86\,GHz data from all three epochs and one additional 43\,GHz image from the observation on 
19 October
2017 closest to the neutrino event. 
This epoch was 
affected by the loss of the Saint Croix (SC) antenna due to hurricane damage and a failure in the Mauna Kea LCP channel. 
As a result, data from this
epoch required further inspection and 
were
excluded from previous publications. We have now calibrated and imaged the data including Mauna Kea (RCP channel only) and all other VLBA antennas 
except
SC.
PKS\,1502+106 was observed with the VLBA at 15, 23 and 43\,GHz on 
30 August 2019. 
PKS\,0215+015 was observed with the VLBA at 15, 23 and 43\,GHz on 
24 March 2022 and in five additional monitoring epochs at the same frequencies 
with a monthly cadence.
All data were reduced using standard methods in \texttt{AIPS} \citep{AIPS} and \texttt{CASA/rPICARD} \citep{CASA,rPICARD} for delay and amplitude calibration. Self-calibration and model fitting 
were
performed using \texttt{DIFMAP} \citep{DIFMAP}. For selected epochs we have also performed polarization calibration using \texttt{PolSolve} \citep{PolSolve}.

\section{Results}
\subsection{TXS\,0506+056 - IC170922A}

\begin{figure}
    \centering
    \includegraphics[width=0.47\columnwidth]{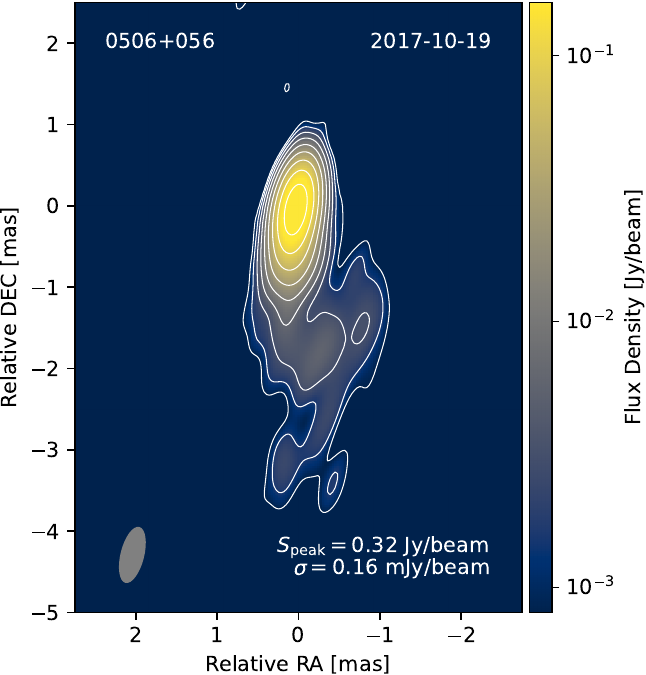}
    \includegraphics[width=0.47\columnwidth]{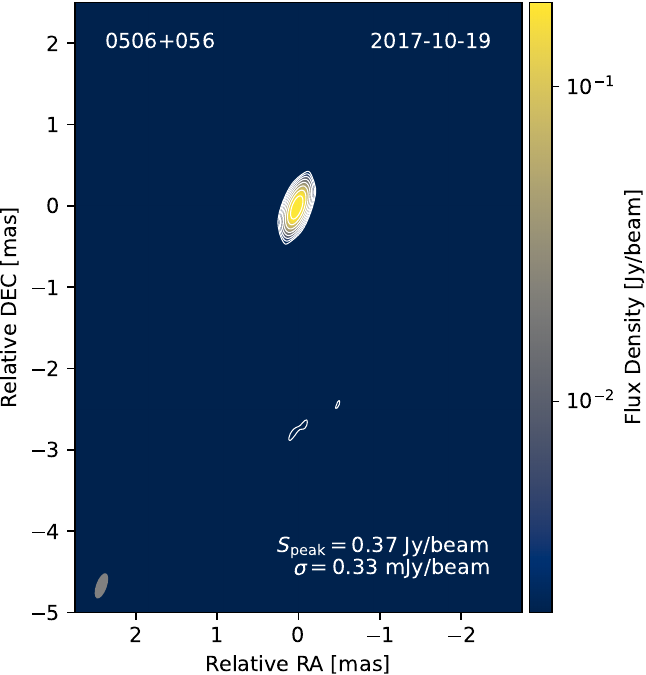}
    \caption{VLBA images of TXS\,0506+056 on Oct 19, 2017 (27 days after neutrino event IC170922A at 43\,GHz (left) and 86\,GHz (right). The lowest contour corresponds to 5\,$\sigma$ in both images, the beam size is indicated by the gray ellipse in the bottom left. The source shows a very compact jet structure
    in all three epochs, especially at 86\,GHz.
    }
    \label{fig:txs0506EpochA}
\end{figure}

We have obtained images of TXS\,0506+056 for the VLBA observation on 
19 October 2017 at 43 and 86\,GHz, only 27 days after neutrino event IC170922A. Both images are shown in Fig.\,\ref{fig:txs0506EpochA}. The 43\,GHz image (left) shows a compact jet structure with a jet pointing to the South. 
Apart from the different beam-shape, the image is 
consistent
with the already published data from 
10 November 2017 \citep{Ros2020}. We have fitted the interferometric visibilities using circular Gaussian components with the \texttt{MODELFIT} algorithm in \texttt{DIFMAP} \citep{DIFMAP}. 
In order to obtain comparable results to the previous study, we 
used the Gaussian model of the 
10 November 2017 epoch, published in \citep{Ros2020}, as a starting model for our Levenberg-Marquardt based model fitting. 
Our model fitting results are shown in Table\,\ref{tab:modelfit_txs0506}. We find that 
by using four
circular Gaussian components the source structure can be modelled without leaving any systematic residuals. 
The sizes and flux densities of individual components agree with the results from the 
10 November 2017 epoch. For the absolute flux density calibration, we estimate an error of $\sim20\,\%$. This means that only C1 seems to have changed significantly, as 
its flux density increased
from 54\,mJy on 
19 October
2017 to 110\,mJy on 
10 November.
However, we like to note that the total flux density of the very compact core region (Core+C1) is in agreement between 
the two
epochs. The component sizes are also largely unaffected between the two epochs. We estimated an uncertainty of $\sim20$\,\% of the FWHM for the size of the components on top of the much smaller formal fit uncertainty. We do not find a significant core expansion between Oct 2017 and Nov 2017 as 
we did
between Nov 2017 and May 2018. This might be because the time between the Oct 2017 and Nov 2017 epoch (22 days) is too short to see such effects. 
Indeed, taking into account the observed core expansion speed found in \citep{Ros2020}, we would expect to measure a core expansion of $\sim5\,\mu$as between the two epochs, which is within the estimated uncertainties. Analogous to the analysis in \citep{Ros2020} we have calculated a brightness temperature for 
each
component following \citep{Kovalev2005}. 
The core component shows a high brightness temperature of $4.8\,\times\,10^{10}$\,K, 
consistent with
a highly relativistic jet. 
This value is in agreement with the later ToO epochs and does not exceed the equipartition limit \citep{Readhead1994}.

The 86\,GHz image from 
19 October 2017 (Fig.\,\ref{fig:txs0506EpochA}, right) shows a very compact core structure and no extended jet emission. 
We have also analyzed the 86\,GHz data from the other two ToO epochs and found no significant hints of extended emission. 
All epochs can be modelled with a single circular Gaussian core component, whose flux densities and sizes are presented in Table\,\ref{tab:modelfit_txs0506}. 
The FWHM of the core in all 86\,GHz components falls below the formal resolution limit $\theta_\mathrm{lim}$, i.e., we use $\theta_\mathrm{lim}$ as an upper limit for the component size to calculate lower limits for the core brightness temperatures at 86\,GHz (c.f., \citep{Kovalev2005}). 
At 86\,GHz the core 
shows
even higher brightness temperatures than at 43\,GHz in all epochs, on the order of $\sim10^{11}$\,K. 
This suggests a highly relativistic environment close to the jet base. 
The lower limits of the brightness temperature do not suggest a significant drop in brightness temperature 
towards May 2018, as suggested by the 43\,GHz data. 
This could be explained by the fact that the core component is unresolved in the 86\,GHz images and only lower limits of the brightness temperature can be determined. 
We expect the absolute flux density error at 86\,GHz to be on the order of $\sim30$\,\%, i.e., we cannot conclude any significant variability of the 86\,GHz core flux density throughout the monitoring period, while the 43\,GHz core component shows a fast rise in radio flux density towards the May 2018 epoch \citep{Ros2020}. 
The radio spectrum of the core component between 43--86\,GHz seems to go from a slightly inverted state close to the neutrino event (spectral index $\alpha\approx 0.4$) to a flat spectrum in the second epoch ($\alpha\approx 0.1$) and is consistent with a steep radio spectrum in May 2018 ($\alpha\approx -0.7$). The inverted spectrum close in time to the neutrino event suggests highly relativistic processes and could provide 
evidence that neutrino production in TXS\,0506+056 
occurs
close to the jet base.

\begin{table}
\centering
\caption{Results of the Gaussian model fitting performed for TXS\,0506+056.}
\label{tab:modelfit_txs0506}
\begin{tabular}{@{}ccccccccc@{}}
\hline
\hline
 Date & Frequency & ID & S & $\Delta\alpha$ & $\Delta\delta$ & FWHM  & $\vartheta_\mathrm{app}$ & T$_\mathrm{b}$ \\
\footnotesize{YYYY-MM-DD} & [GHz] & & [mJy] & [$\mu$as] & [$\mu$as] & [$\mu$as] & [$^\circ$] & [K] \\
\hline
 2017-10-19 & 43 & Core & 284 & - & - & 72 & - & $4.8\,\times\,10^{10}$ \\
  &  & C1 & 54 & -77 & -183 & 105 & 4.8 & $4.3\,\times\,10^{9}$ \\
  &  & C2 & 78 & 15 & -575 & 288 & 12.8 & $8.3\,\times\,10^{8}$ \\
  &  & C3 & 35 & -283 & -1671 & 1118 & 20.1 & $2.5\,\times\,10^{7}$\\
\hline
2017-10-19 & 86 & Core & 369 & - & - & 5 & - & $>2.8\,\times\,10^{11}$  \\ 
2017-11-10 &  & Core & 299 & - & - & 20 & - & $>9.7\,\times\,10^{10}$\\
2018-05-04 &  & Core & 313 & - & - & 15 & - & $>2.7\,\times\,10^{11}$\\
\hline
\multicolumn{9}{@{}l@{}}{\textbf{Notes:} S: total flux density; $\Delta\alpha$ and $\Delta\delta$: component offset from the core in RA and DEC;} \\
\multicolumn{9}{@{}l@{}}{FWHM: Gauss component size; $\vartheta_\mathrm{app}$: apparent opening angle w.r.t. previous feature;}\\
\multicolumn{9}{@{}l@{}}{T$_\mathrm{b}$: brightness temperature.}
\end{tabular}

\end{table}

\subsection{PKS\,1502+106 - IC190730A}

We have calibrated and imaged the 15, 22 and 43\,GHz data of the VLBA ToO observation on 
30 August 2019, 31 days after the associated neutrino event. 
At all frequencies, the source shows a compact, bright core with an extended jet towards the East. 
We are currently facing some issues with the absolute flux calibration of the data and 
have therefore not been able to
determine sensible brightness temperatures for this observation.

\subsection{PKS\,0215+015 - IC220225A}

In a preliminary analysis of the first observation of PKS\,0215+015, 
made on 24 March 2022, 26 days after the associated neutrino event, we have found a very compact jet structure with a high brightness temperature core. The source 
appears to have been in an 
enhanced
polarization state according to single-dish measurements from the Effelsberg 100-m telescope. 
A preliminary polarization calibration of the VLBI data data reveals that the polarized emission originates from the core region. \citep{Eppel2023}

\section{Conclusions \& Outlook}

We have observed TXS\,0506+056, PKS\,1502+106 and PKS\,0215+015, three neutrino candidate blazars, with the VLBA within $\sim1$\,month of their associated neutrino event. In the case of TXS\,0506+056 we have presented the first 86\,GHz VLBI image of the source and one
additional 43\,GHz image closer in time to the neutrino event than in previous publications \citep{Ros2020}. 
We find a very compact jet structure and high-brightness temperature core which suggest a highly relativistic environment close to the jet base. 
Our results are consistent with a previous analysis of two slightly later VLBI observations \citep{Ros2020}. 
However, we do not see a significant core expansion in the 86\,GHz data, since the core is unresolved in all three 86\,GHz epochs.
We are currently 
interpreting 
these high frequency data together with additional 23\,GHz observations, and 15\,GHz data from the MOJAVE database to investigate the jet kinematics in response to the neutrino event. 
The high brightness temperatures and possibly inverted spectrum close in time to the neutrino event in TXS\,0506+056 
may indicate
that neutrino production in TXS\,0506+056 
occurs
close to the jet base.
Additionally, we have observed 
two other neutrino candidate blazars with the VLBA which exhibited major radio flaring during neutrino events IC190730A (PKS\,1502+106) and IC220225A (PKS\,0215+015). 
In the case of PKS\,1502+106 we have found a compact core-jet structure which slightly extends to the East at 15, 22 and 43\,GHz. 
PKS\,0215+015 also shows a very compact jet structure with a high brightness temperature core. 
We also found 
evidence for an enhanced
linear polarization from the core region close in time to the neutrino event \citep{Eppel2023}. 
For both sources we plan to analyze 15\,GHz MOJAVE data to study the jet kinematics to search for new jet components possibly related to the neutrino events. 
Moreover,
we will analyze and compare the polarization maps (linear polarization and EVPA) of all three sources to investigate possible similarities or the existence of a spine-sheath structure \citep{Ghisellini2005}, which can potentially explain neutrino production.

\section*{Acknowledgements}
FE, FR, JH, and MK acknowledge support from Deutsche Forschungsgemeinschaft (DFG, grants 447572188, 434448349, 465409577). PB was supported through a PhD grant from the International Max Planck Research School
(IMPRS) for Astronomy and Astrophysics at the Universities of Bonn and Cologne.
We acknowledge the M2FINDERS project from the European Research Council (ERC) under the European Union’s Horizon 2020 research and innovation programme (grant No 101018682). The National Radio Astronomy Observatory is a facility of the National Science Foundation operated under cooperative agreement by Associated Universities, Inc.

\bibliographystyle{JHEP}
\bibliography{references,mnemonic,jw_abbrv,aa_abbrv}

\end{document}